\documentclass[preprint,showpacs,amsmath,amssymb,floats,nofootinbib]{revtex4}
\usepackage{graphicx}
\usepackage[english]{babel}
\usepackage{amsmath}
\usepackage{amssymb}

\newcommand{\be}{\begin{eqnarray}}
\newcommand{\ee}{\end{eqnarray}}

\newcommand{\dc}{c^{\dagger}}

\newcommand{\nn}{\nonumber}

\def\ket#1{|#1\rangle}
\def\bra#1{\langle #1 |}
\def\ep#1{\langle #1 \rangle}

\begin{document}

\title{The boundary effects of transverse field Ising model}

\author{Yan He}
\email{heyan$_$ctp@scu.edu.cn}
\affiliation{College of Physical Science and Technology,
Sichuan University, Chengdu, Sichuan 610064, China}

\author{Hao Guo}
\affiliation{Department of Physics, Southeast University, Nanjing,Jiangsu 211189, China}

\date{\today}

\begin{abstract}
Advance in quantum simulations using trapped ions or superconducting elements allows detailed analysis of the transverse field Ising model (TFIM), which can exhibit a quantum phase transition and has been a paradigm in exactly solvable quantum systems. The Jordan-Wigner transformation maps the one-dimensional TFIM to a fermion model, but additional complications arise in finite systems and introduce a fermion-number parity constraint when periodic boundary condition (PBC) is imposed. By constructing the free energy and spin correlations with the fermion-number parity constraint and comparing the results to the TFIM with open boundary condition, we show that the boundary effects can become significant for the anti-ferromagnetic TFIM with odd number of sites at low temperature.
\end{abstract}

\maketitle

\section{introduction}
\label{sec:intro}

The quantum transverse field Ising model (TFIM) has been studied for many years because of its relevance to magnetic systems and statistical physics \cite{Lieb,Pfeuty,Sachdev-book,SuzukiBook,DuttaBook}. Quantum simulations of the TFIM using trapped ions~\cite{Kim11,Cui16} or superconducting elements~\cite{Salathe15} allow detailed analyses of static or dynamic properties of the TFIM. The ferromagnetic (FM) or anti-ferromagnetic (AFM) TFIM has a quantum critical point separating the zero-temperature FM or AFM phases and paramagnetic phases in the thermodynamic limit, and influences of the quantum critical point can persist to finite temperatures. Moreover, dynamics of the TFIM across the critical point has been an interesting topic~\cite{Dz,DZ10,DuttaBook,Cui16}.

In one dimension, the TFIM is an exact solvable model and its solution is usually obtained via the Jordon-Wigner transformation \cite{JW,Lieb}, which is a non-local mapping between the spin and fermion operators. Due to the non-local nature of Jordon-Wigner transformation, the TFIM with periodic boundary condition (PBC) cannot be mapped to a free fermion model straightforwardly. The sign of the hopping term of fermions at the periodic boundary depends on whether the total fermion number in the system is even or odd. The resulting fermion Hamiltonian is called the ``a-cycle'' problem in Ref.~\cite{Lieb}. The excitations are not independent of each other because they depend on the parity of the fermion number, which is a global property of the system. In previous treatments~\cite{Pfeuty}, one usually ignores the subtlety of the fermion hopping term at the periodic boundary. The approximation reduces the ``a-cycle'' problem to a genuine free fermion problem, and it is ready to be solved. The error of this approximation can be ignored in the thermodynamic limit when the system becomes infinitely large. Since the phase transition is infinitely sharp at the thermodynamic limit, the approximation has been widely accepted in the literature.

In recent years much progress has been made in manipulating quantum systems, which has made it possible to realize model systems with finite size. For example, quantum simulators using ultracold atoms \cite{Bloch12,Chien15}, trapped ions \cite{Blatt08,Kim11,Blatt12,Cui16}, or superconducting elements \cite{Paraoanu14,Salathe15} have demonstrated interesting quantum phenomena which would have been very difficult to realize in conventional condensed matter systems. Ultracold-atom experiments are usually performed at non-zero temperatures compared to the intrinsic energy scales (such as the quantum degeneracy temperatures) of the model systems. Thus, finite-size and finite-temperature effects can be significant and experimentally accessible. Moreover, the loss of fidelity in operations on trapped ions or superconducting elements \cite{Blatt08} also limits the system size. All those considerations call for a more careful treatment of boundary effects in finite quantum systems.

Here we focus on the detailed calculations of boundary effect of the TFIM with both FM and AFM coupling and  As suggested in Ref. \cite{Lieb1}, the Jordan-Wigner transformation maps the periodic TFIM to a fermion system, and the mapping requires either PBC or anti-periodic boundary condition (APBC) for the fermions depending on the parity of fermion number. Because of this fermion-number parity constraint, applying the Fermi-Dirac distribution to obtain thermodynamic quantities or correlation functions only gives an approximation. To analyze the exact solution, we instead introduce a partition function with alternating sign to evaluate the free energy and spin correlations of the TFIM with PBC. Interestingly, when the fermion-number parity constraint is ignored, the results are almost identical to the TFIM with open boundary condition. We will show that there exist observable differences of magnetization and spin correlations between closed and open boundary TFIM at relative low temperature and finite system size. For FM coupling, this difference is quite small, but for AFM coupling with odd number of sites, this difference is more obvious. The reason is that a ring with odd number of sites is not a bipartite lattice, therefore a classical staggered spin configuration cannot fit in. We can describe this situation as a ``ring frustration'' which was carefully studied in \cite{Li1}. This frustration gives rise to a gapless low energy excitation in the AFM phase in contrast to the gapped excitation in the FM phase. It also causes the obvious difference in spin correlations between closed and open systems.

In the following, we first discuss the exact solution of TFIM of a closed lattice in section \ref{sec:exact}. Then we show how to compute the spin correlations with fermion number parity constraint in section \ref{sec:spin}. In section \ref{sec:open}, we also show the solution of open boundary TFIM. The numerical results and discussions are presented in section \ref{sec:num}.

\section{TFIM with periodic boundary condition}
\label{sec:exact}

The 1D TFIM with PBC is given by
\be
H=-h \sum_{j=1}^N S^z_j-J\sum_{j=1}^N S_j^xS^x_{j+1}
\ee
Here $S^a=\frac{1}{2}\sigma^a$ with $\hbar= 1$, and $\sigma^a$ with $a=x,y,z$ are the three Pauli matrices. We will assume $h\ge 0$, and there are $N$ lattice sites with PBC, so $S^a_{N+1}=S^a_1$. The Jordan-Wigner transformation \cite{JW}
\be
c_n=\exp\Big(\pi i\sum_{j=1}^{n-1}S_j^+S_j^-\Big)S_n^-\qquad
\dc_n=\exp\Big(-\pi i\sum_{j=1}^{n-1}S_j^+S_j^-\Big)S_n^+
\ee
Here $S_j^{\pm}=S_j^x\pm iS_j^{y}$. Then the Hamiltonian becomes
\be
H=\frac{hN}{2}-h\sum_{i=1}^N\dc_i c_i
-\frac{J}{4}\sum_{i=1}^{N-1}(\dc_i-c_i)(\dc_{i+1}+c_{i+1})
+\frac{J}{4}\exp(i\pi N_f)(\dc_N-c_N)(\dc_{1}+c_{1})
\ee
In the fermion Hamiltonian, $N_f=\sum_{j=1}^{N}\dc_jc_j$ is the total fermion number related to $S^z$ via $\dc_jc_j=S^z_j+\frac12$. Since $S^z_j$ does not commute with the Hamiltonian, the total fermion number $N_f$ is not conserved in the fermion model. Importantly, the presence of the last term may cause the resulting fermion Hamiltonian not to follow PBC. In previous works the factor $\exp(i\pi N_f)$ was ignored and the fermion Hamiltonian follows PBC again~\cite{Pfeuty}. The error from the approximation becomes negligible after the thermodynamical limit has been taken since one term only produces a correction of order $1/N$.

Nevertheless, we will show that finite-size effects reflecting subtle boundary effects indeed have observable consequences by taking a more careful treatment of the fermion Hamiltonian. Firstly, the fermion Hamiltonian can be cast into a form similar to the BCS Hamiltonian of conventional superconductors \cite{MahanBook}. Explicitly,
\begin{eqnarray}\label{Ham}
H&=&\sum_{j=1}^N\left[-\frac J4(\dc_j c_{j+1}+\dc_{j+1} c_j)-h \dc_j c_j
+\frac J4 c_j c_{j+1}+\frac J4 \dc_{j+1} \dc_{j}\right]+\frac{hN}{2}
\end{eqnarray}
with the following boundary conditions
\be
&&c_{N+1}=c_1,\quad\mathrm{for}\quad N_f\equiv1(mod\,2), \nonumber \\
&&c_{N+1}=-c_1\quad\mathrm{for}\quad N_f\equiv0(mod\,2)
\ee
Thus, PBC or APBC is imposed on the case with odd or even total fermion number, respectively.

Different from Ref.~\cite{Lieb}, here we first transform the Hamiltonian to momentum space by introducing
\be
c_n=\frac1{\sqrt{N}}\sum_k c_k e^{ikn},\quad\dc_n=\frac1{\sqrt{N}}\sum_k \dc_k e^{-ikn}
\ee
where $j$ labels the site and $k$ labels the momentum. The boundary conditions for the fermions then require
\be
&&k\in\Lambda_a,\, \Lambda_a=\Big\{\pm\frac{\pi}{N},\,\pm\frac{3\pi}{N},\cdots,\pm\frac{(N-1)\pi}{N}\Big\}
\quad\mbox{for APBC}\nn\\
&&k\in\Lambda_p,\, \Lambda_p=\Big\{0,\,\pm\frac{2\pi}{N},\,\pm\frac{4\pi}{N},\cdots,\pm\frac{(N-2)\pi}{N},\,\pi\Big\}
\quad\mbox{for PBC}\nn
\ee
when the total number of sites $N$ is even. We will refer to the first case as the APBC channel and the second as the PBC channel. One has to consider the contributions from both channels and include the correct $k$ values from the set $\Lambda_a$ or $\Lambda_p$. The difference of $\Lambda_a$ and $\Lambda_p$ is negligible in the limit $N\to\infty$, so there is no need to distinguish these two channels in the thermodynamic limit. This subtlety of performing the Fourier transform was also mentioned in Ref.~\cite{DuttaBook}.

The total number of lattice sites $N$ is chosen as an even number for the FM case. For AFM case, we take $N$ as an odd number in order to investigate the effects of ring frustration. In this case, $k$ takes the following values
\be
&&k\in\Lambda_a,\, \Lambda_a=\Big\{\pm\frac{\pi}{N},\,\pm\frac{3\pi}{N},\cdots,\pm\frac{(N-2)\pi}{N},\,\pi\Big\}
\quad\mbox{for APBC}\nn\\
&&k\in\Lambda_p,\, \Lambda_p=\Big\{0,\,\pm\frac{2\pi}{N},\,\pm\frac{4\pi}{N},\cdots,\pm\frac{(N-1)\pi}{N}\Big\}
\quad\mbox{for PBC}\nn
\ee
Note that $k=\pi$ is moved from PBC channel to APBC channel. In the following, we only show the derivation of the FM case. The results of AFM case are similar.

For the FM case, the Hamiltonians in APBC and PBC channels are
\be
&&H_a=\sum_{k\in\Lambda_a'}\Big[\xi_k\dc_kc_k+\xi_k\dc_{-k}c_{-k}
+i\frac J2\sin k c_{-k}c_{k} -i\frac J2\sin k\dc_k\dc_{-k}\Big]+\frac{hN}{2}\nn\\
&&H_p=\sum_{k\in\Lambda_p'}\Big[\xi_k\dc_kc_k+\xi_k\dc_{-k}c_{-k}
+i\frac J2\sin k c_{-k}c_{k} -i\frac J2\sin k\dc_k\dc_{-k}\Big]+\frac{hN}{2}\nn\\
&&\qquad+\xi_0\dc_0c_0+\xi_{\pi}\dc_{\pi}c_{\pi}
\ee
here $\xi_k=-\frac J2\cos k-h$ and $k$ is taken values from the following two sets, $\Lambda_a'=\{k|k\in\Lambda_e,\,k>0\}$, $\Lambda_p'=\{k|k\in\Lambda_o,\,k>0,\,k\neq\pi\}$.

The Hamiltonian can be diagonalized by a Bogoliubov transformation
\be
\left(
  \begin{array}{c}
    c_k \\
    \dc_{-k}
  \end{array}
\right)=\left(
  \begin{array}{cc}
    u_k & v_k \\
    -v_k^* & u_k
  \end{array}
\right)\left(
  \begin{array}{c}
    \eta_k \\
    \eta^{\dagger}_{-k}
  \end{array} \right)
\ee
Here we choose
\be
u_k=\sqrt{\frac{E_k+\xi_k}{2E_k}},\quad v_k=i\,\mbox{sgn}k \sqrt{\frac{E_k-\xi_k}{2E_k}},
\label{uvk}
\ee
and $\mathrm{sgn}k$ is the sign of $k$. The quasi-particle dispersion is
\be
E_k=\sqrt{(J/2)^2+h^2+Jh\cos k}.\label{eq:Ek}
\ee
The quasi-particle dispersion can become gapless when $h=J/2$, which is the quantum critical point of the TFIM separating the FM or AFM phases from paramagnetic phases in the thermodynamic limit \cite{Kogut,Sachdev-book}. For finite-size systems, thermodynamical quantities have no singular behaviors around this quantum critical point.

The diagonalized Hamiltonian in the APBC and PBC channels can be written as
\be
&&H_a=E_0+\sum_{k\in\Lambda_a'}E_k(\eta_k^{\dagger}\eta_k+\eta_{-k}^{\dagger}\eta_{-k})\nn\\
&&H_p=E_1+\sum_{k\in\Lambda_p'}E_k(\eta_k^{\dagger}\eta_k+\eta_{-k}^{\dagger}\eta_{-k})\nn\\
&&\qquad+\xi_0\dc_0c_0+\xi_{\pi}\dc_{\pi}c_{\pi}
\ee
with $E_0=-\sum_{k\in\Lambda_a'}E_k$ and $E_1=h-\sum_{k\in\Lambda_p'}E_k$.
Special care should be taken for the eigenmodes with $k=0$ and $\pi$ since at $k=0,\pi$ the energy gap is zero, so there is no need for the Bogoliubov transformation.

Before we discuss how to calculate statistical average, let us briefly discuss the ground state and low energy modes of TFIM with both FM and AFM coupling. In the TFIM with FM coupling, one can numerically check that the ground state of the system is the state with no quasi-particle. Therefore one should impose APBC and the ground state is $\ket{0}_a$ satisfying $\eta_k\ket{0}_a=0$ for all $k\in\Lambda_a$. The low energy excitations are states like $\eta_k^{\dagger}\ket{0}_a$ which is clearly gapped. In the TFIM with AFM coupling with odd number of sites, we can introduce a similar state $\ket{0}'_p$ satisfying $\eta_k\ket{0}'_p=0$ for all nonzero $k\in\Lambda_p$ and $c_0\ket{0}'_p=0$. Then the ground state in AFM case is $\dc_0\ket{0}'_p$. Furthermore, the energy of excited state $\eta_k\ket{0}'_p$ approaches to the ground state energy as $k\to 0$ in the thermodynamic limit. Therefore in the AFM case, the low energy states are gapless in contrast to the FM case \cite{Li1}. Therefore we expect that the boundary effects are more obvious in the AFM case with odd number of sites.

\section{Free energy and spin correlations}
\label{sec:spin}

At finite $T$, one should not use the Fermi distribution to compute the statistical average because APBC and BPC channels have to be treated differently. In order to compute the statistical average, we first list the whole spectra as
\be
E(\{n_k\})=\sum_{k\in\Lambda_a}\Big[-\frac{E_k}{2}(1-n_k)+\frac{E_k}{2}n_k\Big],\quad \sum_k n_k\equiv\nu_a(mod\,2)\\
E(\{n_k\})=\sum_{k\in\Lambda_p}\Big[-\frac{E_k}{2}(1-n_k)+\frac{E_k}{2}n_k\Big],\quad \sum_k n_k\equiv\nu_p(mod\,2)
\ee
Here $n_k=0$ or $1$ is the occupation number of eigenmode $\eta_k$. In order to treat all eigenmode in the same fashion, we also apply the
Bogoliubov transformation to $c_0$ and $c_{\pi}$ separately to obtain eigenmodes $\eta_0$ and $\eta_{\pi}$. Since Bogoliubov transformation of a pair of fermions will not change the fermion number parity, one would naively expect that $\sum_k n_k$ is even for APBC and is odd for PBC.  But the eigenmode of  $\eta_0$ and $\eta_{\pi}$ are special. If its energy is already positive, then the Bogoliubov transformation did nothing. On the other hand, if its energy is negative, the Bogoliubov transformation will switch the particle and hole and in turn change the total fermion number by one and flip the fermion number parity.

For the FM case, both $\eta_0$ and $\eta_{\pi}$ appear in the PBC channel. Thus we do have even $\sum_k n_k$ for APBC channel as naively expected. In the APBC channel, the dispersion of $\eta_0$ is $\xi_0=-J/2-h$ which is always negative. Thus the fermion number parity is always flipped once due to the Bogoliubov transformation of $\eta_0$. For $\eta_{\pi}$ we have  $\xi_{\pi}=J/2-h$, thus for $h>J/2$ the fermion number will be flipped again, but for $h<J/2$, this will not happen. In summary, we find that
\be
\nu_a=0,\quad
\nu_p=\left\{
  \begin{array}{c}
    0,\quad h<J/2 \\
    1,\quad h>J/2
  \end{array}
\right.\label{nua}
\ee
Similarly, for the AFM case with odd number of sites, $\eta_0$ is in the PBC channel and $\eta_{\pi}$ is in the APBC channel. Since $J<0$, we have $\xi_{\pi}=J/2-h<0$ and the fermion number parity is always flipped once in the APBC channel. In the PBC channel, $\xi_0=-J/2-h$ is negative for for $h>|J|/2$ and the fermion number will be flipped in this case, but for $h<|J|/2$ this will not happen. In summary, we find that
\be
\nu_a=1,\quad
\nu_p=\left\{
  \begin{array}{c}
    1,\quad h<|J|/2 \\
    0,\quad h>|J|/2
  \end{array}
\right.\label{nup}
\ee

Now we have the whole spectra of the system. It is straightforward to compute the partition function. Each un-occupied eigenmode $\ket{n_k=0}$ has energy $-E_k/2$. Each occupied eigenmode $\ket{n_k=1}$ has energy $E_k/2$. The total number eigenmodes is the system size $N$. The
partition functions $Z=\sum_{\{n_k\}}e^{-E(\{n_k\})/T}$ for APBC and PBC channels are
\be
Z_{a1}=\prod_{k\in\Lambda_a}\Big(e^{\frac{E_k}{2T}}+e^{-\frac{E_k}{2T}}\Big),\quad
Z_{p1}=\prod_{k\in\Lambda_p}\Big(e^{\frac{E_k}{2T}}+e^{-\frac{E_k}{2T}}\Big)
\ee
In order to distinguish the even and odd number of excitations, we also introduce the partition function with alternating sign $Z=\sum_{\{n_k\}}(-1)^{\sum_kn_k}e^{-E(\{n_k\})/T}$. For APBC and PBC channels, the results are.
\be
Z_{a2}=\prod_{k\in\Lambda_a}\Big(e^{\frac{E_k}{2T}}-e^{-\frac{E_k}{2T}}\Big),\quad
Z_{p2}=\prod_{k\in\Lambda_p}\Big(e^{\frac{E_k}{2T}}-e^{-\frac{E_k}{2T}}\Big)
\ee
According to the fermion number parity given by Eq.(\ref{nua}) and (\ref{nup}), we find the total partition function for the FM case and AFM case as
\be
&&Z_{FM}=\frac12\Big[Z_{a1}+Z_{a2}+Z_{p1}-\mbox{sgn}(h-\frac J2)Z_{p2}\Big]\\
&&Z_{AFM}=\frac12\Big[Z_{a1}-Z_{a2}+Z_{p1}+\mbox{sgn}(h-\frac{|J|}2)Z_{p2}\Big]
\ee
Then the free energy per particle for both cases are $F_{FM}/N=-\frac{T}N\ln Z_{FM}$ and  $F_{AFM}/N=-\frac{T}N\ln Z_{AFM}$. We can verify that the above free energy will reduce the correct ground state energy as $T\to0$. In the FM case and the low $T$ limit, we have
\be
Z_{a1}=Z_{a2}\approx\prod_{k\in\Lambda_a}e^{\frac{E_k}{2T}},\quad
Z_{p1}=Z_{p2}\approx\prod_{k\in\Lambda_p}e^{\frac{E_k}{2T}}
\ee
One can verify that $Z_{a1}>Z_{p1}$ in the FM case, thus the ground state energy of FM case
is $E_{FM}=-T\ln Z_{a1}=\frac12\sum_{k\in\Lambda_a}E_k$ and the ground state is $\ket{0}_a$ annihilated by all $\eta_k$ as discussed in section \ref{sec:intro}.

In the AFM case and the low $T$ limit, we find
\be
&&Z_{a1}-Z_{a2}\approx2\prod_{k\in\Lambda_a,k\neq k_1}\exp(\frac{E_k}{2T})\cdot\exp(-\frac{E_{k1}}{2T}),\quad
k_1=\pi/N\\
&&Z_{p1}-Z_{p2}\approx2\prod_{k\in\Lambda_a,k\neq 0}\exp(\frac{E_k}{2T})\cdot\exp(-\frac{E_0}{2T}),\quad
h<|J|/2\\
&&Z_{p1}+Z_{p2}\approx2\prod_{k\in\Lambda_p}\exp(\frac{E_k}{2T}),\quad h>|J|/2
\ee
One verify that the lower two lines of above equations are always larger in the AFM case. Therefore the ground state energy is $E_{AFM}=\frac12\sum_{k\in\Lambda_p}E_k+E_0$ for $h<|J|/2$ and $E_{AFM}=\frac12\sum_{k\in\Lambda_p}E_k$ for $h>|J|/2$. Since there is a particle-hole exchange for $c_0$ when $h>|J|/2$, one can see the ground state is just $\dc_0\ket{0}'_p$ as discussed in section \ref{sec:intro}.

The spin correlations, such as $C_x(n)=\ep{S_i^x S_{i+n}^x}$, $C_y(n)=\ep{S_i^y S_{i+n}^y}$, and $C_z(n)=\ep{S_i^z S_{i+n}^z}$, can be obtained from the correlation functions of the fermion operators. Following Ref.~\cite{Lieb}, we introduce the operators $A_i=\dc_i+c_i$ and $B_i=\dc_i-c_i$. Then the correlation functions of $A_i$ and $B_i$ are
\be
&&\ep{A_i A_j}=-\ep{B_i B_j}=-\delta_{ij}\nn\\
&&\ep{B_i A_j}=-\ep{A_j B_i}=G(i-j)\label{eq:Gn}
\ee
Here the function $G(i-j=n)$ is still to be determined. Then $C_x(n)$, $C_y(n)$, and $C_z(n)$ can be expressed in terms of $G(n)$ as discussed in Ref.~\cite{Pfeuty}. Explicitly,
\be\label{eq:Cxyz}
C_x(n)&=&\frac14\det\left(
                           \begin{array}{cccc}
                             G(-1) & G(-2) & \cdots & G(-n) \\
                             G(0) & G(-1) & \cdots & G(-n+1) \\
                             \vdots & \vdots & \vdots & \vdots \\
                             G(n-2) & G(n-3) & \cdots & G(-1)
                           \end{array}
                         \right), \nonumber \\
C_y(n)&=&\frac14\det\left(
                           \begin{array}{cccc}
                             G(1) & G(0) & \cdots & G(2-n) \\
                             G(2) & G(1) & \cdots & G(3-n) \\
                             \vdots & \vdots & \vdots & \vdots \\
                             G(n) & G(n-1) & \cdots & G(1)
                           \end{array}
                         \right), \nonumber \\
C_z(n)&=&-\frac14G(n)G(-n).
\ee
Moreover, the $z$-direction magnetization $M_z=\ep{S^z_i}$ can be expressed as
\be\label{eq:Mz}
M_z=\frac12 G(0).
\ee

Now we focus on the evaluation of $G(n)$. It can be expressed as
\be
G(n)=\ep{\dc_n\dc_0}+\ep{\dc_n c_0}-\ep{c_n\dc_0}-\ep{c_nc_0}
\ee
Before computing the statistical average, we first compute the expectation of these operators with occupied or un-occupied eigenmodes.
Explicitly, the expectations can be obtained as follows.
\be
&&\bra{0_k}c_n c_0\ket{0_k}=-\bra{1_k}c_n c_0\ket{1_k}=-e^{ikn}u_k v_{k}\nn\\
&&\bra{0_k}c_n \dc_0\ket{0_k}=-\bra{1_k}c_n \dc_0\ket{1_k}=e^{ikn}u_k^2\nn\\
&&\bra{0_k}\dc_n c_0\ket{0_k}=-\bra{1_k}\dc_n c_0\ket{1_k}=e^{ikn}|v_k|^2\nn\\
&&\bra{0_k}\dc_n \dc_0\ket{0_k}=-\bra{1_k}\dc_n \dc_0\ket{1_k}=e^{ikn}u_k v_{k}\nn
\ee
Here $\ket{0_k}=\ket{n_k=0}$ is an un-occupied eigenmode and $\ket{1_k}=\ket{n_k=1}$ is an occupied eigenmode.  The coefficients $u_k$ and $v_k$ are defined in Eq. (\ref{uvk}). Assemble the above results, we find that that when computing $G(n)$, each un-occupied eigenmode $\ket{n_k=0}$ contribute $\mathcal{G}(n,k)$ and each occupied eigenmode $\ket{n_k=1}$ contribute $-\mathcal{G}(n,k)$, where $\mathcal{G}(n,k)$ is given by
\be
\mathcal{G}(n,k)=\frac{1}{E_k}\Big[h \cos kn+\frac J2 \cos k(n+1)\Big]
\ee

Then following similar steps as in computing the partition function, we find that $G(n)$ for FM and AFM cases are given as follows.
\be
&&G_{FM}(n)=\frac1N\Bigg[\frac{Z_{a1}}{Z_{FM}}\sum_{k\in\Lambda_a}\mathcal{G}(n,k)\tanh\frac{E_k}{2T}
+\frac{Z_{a2}}{Z_{FM}}\sum_{k\in\Lambda_a}\mathcal{G}(n,k)\coth\frac{E_k}{2T}\nn\\
&&\quad+\frac{Z_{p1}}{Z_{FM}}\sum_{k\in\Lambda_p}\mathcal{G}(n,k)\tanh\frac{E_k}{2T}
-\mbox{sgn}(h-\frac J2)\frac{Z_{p2}}{Z_{FM}}\sum_{k\in\Lambda_p}\mathcal{G}(n,k)\coth\frac{E_k}{2T}\Bigg]\\
&&G_{AFM}(n)=\frac1N\Bigg[\frac{Z_{a1}}{Z_{AFM}}\sum_{k\in\Lambda_a}\mathcal{G}(n,k)\tanh\frac{E_k}{2T}
-\frac{Z_{a2}}{Z_{AFM}}\sum_{k\in\Lambda_a}\mathcal{G}(n,k)\coth\frac{E_k}{2T}\nn\\
&&\quad+\frac{Z_{p1}}{Z_{AFM}}\sum_{k\in\Lambda_p}\mathcal{G}(n,k)\tanh\frac{E_k}{2T}
+\mbox{sgn}(h-\frac{|J|}2)\frac{Z_{p2}}{Z_{AFM}}\sum_{k\in\Lambda_p}\mathcal{G}(n,k)\coth\frac{E_k}{2T}\Bigg]
\ee
Now the spin correlations of the TFIM can be calculated exactly with the APBC and PBC channels considered separately.

\section{TFIM with open boundary condition}
\label{sec:open}

For the TFIM with open boundary condition, there is no translational invariance, so transforming the Hamiltonian to momentum space does not lead to further simplification. We will closely follow the method of Ref.~\cite{Lieb} and implement a Bogoliubov transformation in real space. The Hamiltonian is cast in the form
\be
H=\sum_{i,j}\Big[\dc_i A_{ij} c_j+\frac12(\dc_i B_{ij}\dc_j-c_i B_{ij}c_j)\Big].
\ee
Here $A$ is a Hermitian matrix and $B$ is an anti-symmetric matrix. For the TFIM,
$A_{ij}=-h\delta_{ij}-\frac J4(\delta_{i,j+1}+\delta_{i+1,j})$ and
$B_{ij}=\frac J4(\delta_{i,j+1}-\delta_{i+1,j})$.
Next, we introduce the quasi-particle annihilation and creation operators as
\be
&&\eta_k=\sum_i(g_{ki}c_i+h_{ki}\dc_i),\quad\eta_k^{\dagger}=\sum_i(g_{ki}\dc_i+h_{ki}c_i).
\ee
Note that $k$ is an integer index not related to the momentum. We also define (with $i$ being the site index) $\phi_{ki}=g_{ki}+h_{ki}$ and $\psi_{ki}=g_{ki}-h_{ki}$, which may be considered as the wave functions of the quasi-particles labeled by index $k$. By requiring that $\eta_k$ diagonalizes the Hamiltonian, $\phi_{ki}$ and $\psi_{ki}$ should satisfy
\be
\sum_j(A+B)_{ij}\phi_{kj}=\lambda_k\psi_{ki},\quad\sum_j(A-B)_{ij}\psi_{kj}=\lambda_k\phi_{ki}.
\ee
Therefore, we can solve $\phi_{k}$ and $\psi_{k}$ from the eigenvalue equations
\be
[(A-B)(A+B)]_{ij}\phi_{kj}=\lambda_k^2\phi_{ki},\quad
[(A+B)(A-B)]_{ij}\psi_{kj}=\lambda_k^2\psi_{ki}
\ee
We can choose $\phi_{ki}$ to be real and satisfy $\sum_k\phi_{ki}\phi_{kj}=\delta_{ij}$. The same is true for $\psi_{ik}$. Here $\lambda_k$ is the energy spectrum.

In order to compute the spin correlations, we introduce operators $A_i=\dc_i+c_i$ and $B_i=\dc_i-c_i$ again, and one obtains
$A_i=\sum_k(\phi_{ki}\eta_k+\phi_{ki}\eta_k^{\dagger})$,
$B_i=\sum_k(\psi_{ki}\eta_k^{\dagger}-\psi_{ki}\eta_k)$.
Their correlation functions are given by
\be\label{Gd}
&&\ep{A_i A_j}=\delta_{ij},\quad \ep{B_i B_j}=-\delta_{ij},\\
&&\ep{B_i A_j}=-\ep{A_j B_i}=G(i,j)=-\sum_k\phi_{ki}\psi_{kj}\tanh\frac{\lambda_k}{2T}. \nonumber
\ee

For the open-boundary TFIM, the matrices $(A-B)(A+B)$ and $(A+B)(A-B)$ are tri-diagonal, so they can be diagonalized as follows. The eigenvectors are given by
\be
&&\phi_{k}=N_k(\sin\theta_k,\,\sin2\theta_k\cdots,\sin N\theta_k)^t, \nonumber \\
&&\psi_{k}=N_k(\sin N\theta_k,\,\sin(N-1)\theta_k,\cdots,\sin\theta_k)^t.
\ee
Here $N_k=(\sum_n\sin n\theta_k)^{-1/2}$ is a normalization factor, the superscript $t$ denotes the transpose, and $\theta_k$ with $k=1,\cdots,N$ are the $N$ roots of the equation
\be
\frac{\sin(N+1)\theta}{\sin N\theta}=-\frac{J}{2h}
\ee
The corresponding eigenvalues are
\be
\lambda_k=\sqrt{(J/2)^2+h^2+Jh\cos\theta_k}
\ee
which have the same form as $E_k$ in Eq.(\ref{eq:Ek}), but $\theta_k$ is different from the momentum. We remark that the energy gap does not fully close at the critical point $h=J/2$ until the system reaches the thermodynamic limit.

After obtaining $\phi_k$ and $\psi_k$, it is straightforward to find $G(i,j)$ by Eq.~(\ref{Gd}) and obtain the spin correlations and magnetization via Eqs.~\eqref{eq:Cxyz} and \eqref{eq:Mz}.
With the energy spectrum, the free energy density is given by
\be
\frac FN=-\frac TN\sum_k\ln\Big(2\cosh\frac{\lambda_k}{2T}\Big).
\ee

\section{Results and discussions}
\label{sec:num}

\begin{figure}[t]
\includegraphics[clip,width=0.8\textwidth]{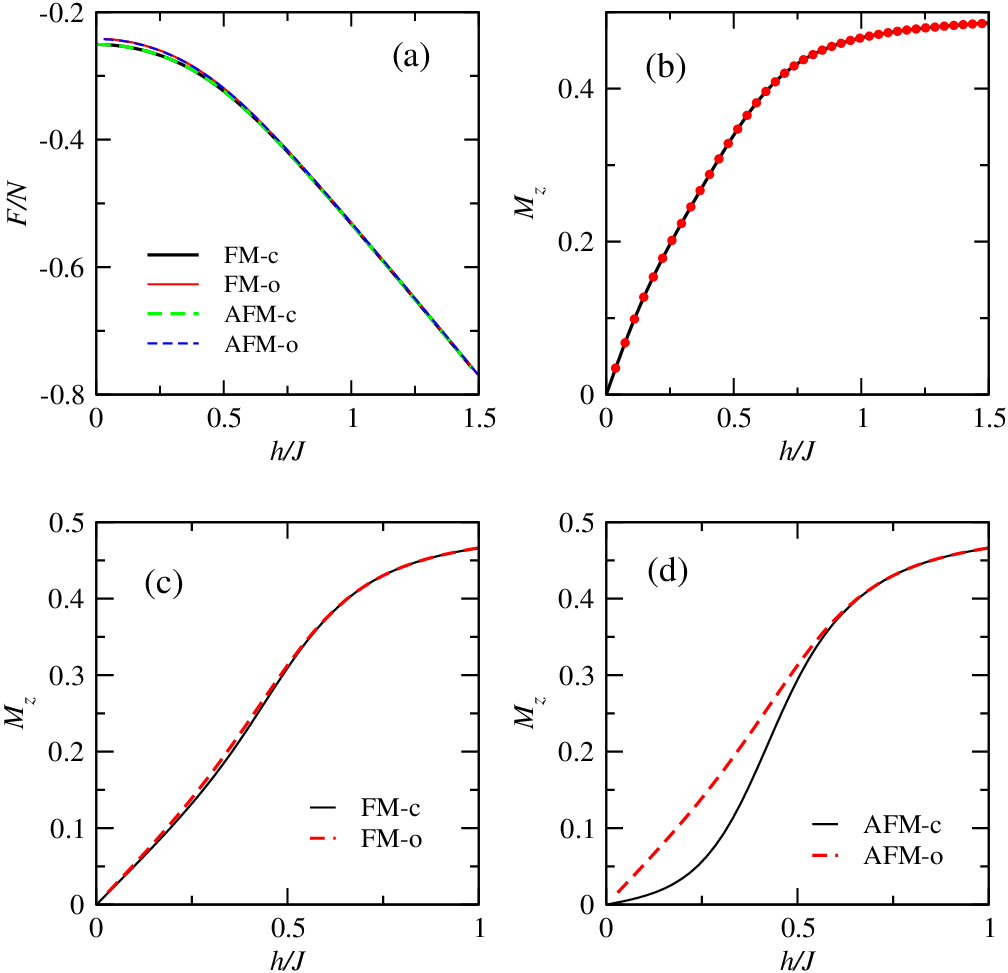}
\caption{(a) The free energy per particle of the TFIM as a function of $h/J$. The four different curves corresponds to the FM coupling with closed and open boundary, AFM coupling with closed and open boundary. (b) shows the $z$-direction magnetization $M_z$ as function of $h/J$ for the AFM coupling. The black curve is obtained by analytic formula, the red dots are obtained by exact diagonalization. (c) show $M_z$ of as a function of $h/J$ for the FM coupling with both closed boundary (black) and open boundary (red dashed). (d) show $M_z$ of as a function of $h/J$ for the AFM coupling with both closed boundary (black) and open boundary (red dashed).}\label{Mz}
\end{figure}

\begin{figure}[t]
\centerline{\includegraphics[clip,width=0.9\textwidth]{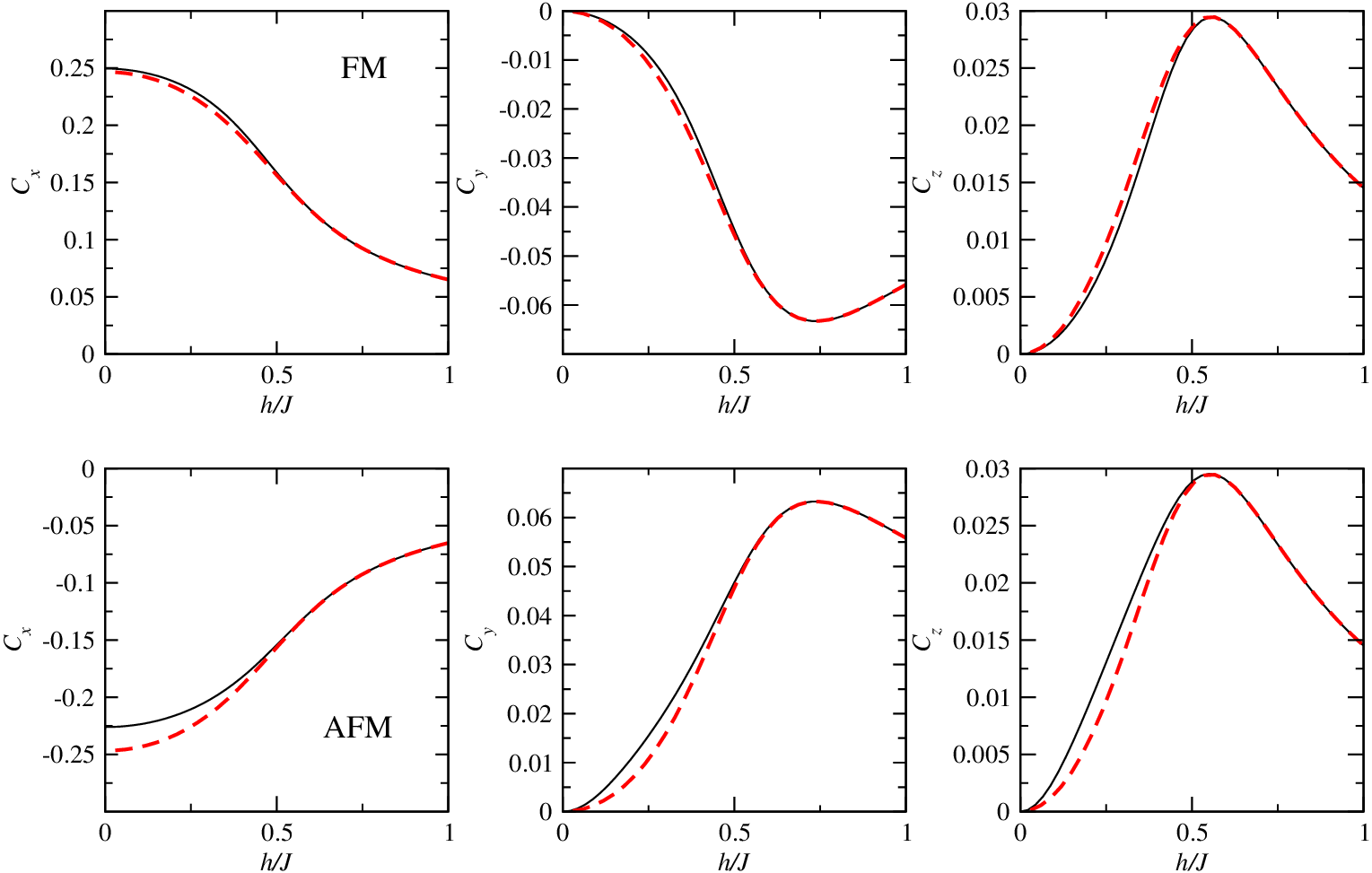}}
\caption{(a), (b), and (c) show the nearest-neighbor spin correlation functions $C_x(1)$, $C_y(1)$, $C_z(1)$ as functions of $h/J$ for the TFIM with FM coupling. (d), (e), and (f) show $C_x(1)$, $C_y(1)$, $C_z(1)$ as functions of $h/J$ for the TFIM with AFM coupling.
The black and red dashed curves correspond to closed and open boundary system.}\label{C-h}
\end{figure}

Here we present numerical results of the TFIM of both FM and AFM couplings. We will show that for a finite size TFIM at low temperature, there is an appreciable difference between closed and open boundary system. For the FM coupling, most results are obtained with $N=20$ lattice sites. For the AFM coupling, we only consider the system with odd number of sites such as $N=21$, which has a ring frustration.  The same conclusion holds for other moderate sized finite systems. On the other hand, for very large system and very high $T$, the closed and open boundary system will give almost the same results. Here the magnitude of the coupling $|J|$ is taken as an energy unit. For convenience we denote $|J|$ as $J$ in this section, that is, ignore the minus sign of $J$ in the AFM case.

Figure \ref{Mz} shows the free energy $F$ and $z$-direction magnetization $M_z$ of the TFIM as a function of $h/J$ at temperature $T/J=0.1$. In panel (a), the black, red, green dashed and blue dashed curves represent the free energy corresponding to the FM coupling with closed and open boundary, AFM coupling with closed and open boundary respectively. There is almost no observable difference in $F$ at this temperature for these for different choices of coupling and boundary conditions. The panel (c) show $M_z$ of as a function of $h/J$ for the FM coupling with both closed boundary (black) and open boundary (red dashed). One can see some very small difference these two curves in the FM phase when $h/J<0.5$. The panel (d) show $M_z$ of as a function of $h/J$ for the AFM coupling with both closed boundary (black) and open boundary (red dashed). One can see the closed boundary results are obvious below the open boundary results in the AFM phase when $h/J<0.5$. In the ring frustration case, the low energy excitation in the AFM phase is gapless in contrast to the gapped excitation in the FM phase. At moderate low $T$, these low energy excitations makes very important contribution to the statistical average. This gives rise to the difference observed in panel (d). In order to build some confidence of the analytical formula we derived last section, we compare the analytical results (black curve) with the results obtained by exact diagonalization (red dots) in panel (b). Here we compare $M_z$ of system with AFM coupling and sites number $N=9$. One can see they agree with each other perfectly.

In Figure \ref{C-h}, we show the nearest-neighbor spin correlations $C_x(1)$, $C_y(1)$, $C_z(1)$ as functions of $h/J$. The upper three panels are the spin correlations for the TFIM with FM coupling. The lower three panels are the spin correlations for the TFIM with AFM coupling. The system size is $N=20$ for the FM case, and $N=21$ for the AFM case. We also assume $T/J=0.1$ as before. It is clear that in the upper panels, there are slightly difference between the closed and open boundary system in the FM phase when $h/J<0.5$. The difference become more dramatic in the lower panels, especially for $C_x(1)$. Again this is due to the appears of gapless excitations in the AFM phase. For $C_x(1)$, the difference between closed and open boundary system even persist to $h=0$. When there is no external field, the low energy gapless modes become degenerate with the ground state \cite{Li1}, which makes the difference of $C_x(1)$. Note that in the AFM case, the neighboring spins are mostly opposite to each other. Therefore $C_x(1)$ is negative for AFM case, while it is positive for the FM case.

\begin{figure}[t]
\centerline{\includegraphics[clip,width=0.8\textwidth]{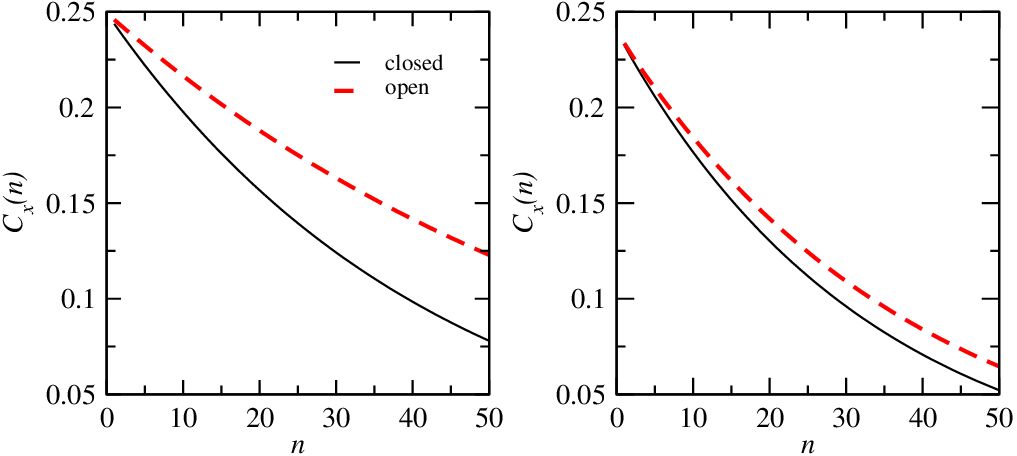}}
\caption{The $x$-direction spin correlation functions $C_x(n)$, as a function of the separation $n$ for the TFIM with AFM coupling at $T/J=0.1$. The left panel is for $h/J=0.05$. The right panel is for $h/J=0.2$. The black and red dashed curves are results of closed and open boundary systems respectively.}\label{Cn}
\end{figure}

In Figure \ref{Cn}, we show the $x$-direction spin correlation functions $C_x(n)$, as a function of the distance $n$ for the TFIM with AFM coupling at $T/J=0.1$. Comparing to the previous figures, here we take a much larger system size $N=101$ and we plot $C_x(n)$ up to $n=50$. For the left and right panels, we take the external field as $h/J=0.05$ and $h/J=0.2$ respectively. For such a large sized system, the short ranged spin correlation $C_x(1)$ are the same for both closed and open boundary system. But the long range behavior of $C_x$ is quite different for closed and open systems. For small $h$,  $C_x(n)$ of the closed system decays faster than the open system. This difference become much smaller when $h$ is approaching the critical value $h/J=0.5$. Again, one can expect that the difference for small $h$ is due to the gapless modes which become almost degenerate with the ground state at small $h$.

In summary, although the boundary effects of the TFIM may be neglected in the thermodynamic limit, we find that there are still some observable differences in the finite size system with low $T$. While these differences in $M_z$ and spin correlations are quite small for the TFIM with FM coupling, they become more significant for TFIM with AFM coupling and odd number of sites. In this case, the system has a ring frustration which makes the low energy excitations gapless. These gapless modes give rise to different behaviors of the spin correlations which are even persistent to quite large system as shown in Figure \ref{Cn}. In deriving these results, we found a way to compute the statistical average with fermion number parity constraint, which may be useful for other applications.

\textit{Acknowledgment ---}
We thank Peng Li and Chih-chun Chien for stimulating discussions. Yan He is supported by NSFC under grant No. 11404228. Hao Guo is supported by NSFC under grant No. 11204032.


%

\end{document}